\documentclass[12pt]{article}
\begin{document}

\begin{flushright}
ITEP/TH-49/99\\
hep-th/9909...
\end{flushright}
\vspace{0.5cm}
\begin{center}

{\LARGE\bf On flat connections with non-diagonalizable holonomies}
\date{today}

\bigskip{{\Large K.Selivanov } \\
ITEP, Moscow, 117259, B.Cheryomushkinskaya 25}
\end{center}
\bigskip

\begin{abstract}
Recently the long-standing puzzle about counting the Witten 
index in $N=1$ supersymmetric gauge theories was resolved. 
The resolution was based on existence
(for higher orthogonal $SO(N),\;  N \geq 7$ and exceptional 
gauge groups) of flat connections on $T^{3}$ which have 
commuting holonomies but cannot be gauged to a Cartan torus.
A number of papers
has been  published which studied moduli spaces and some topological
characteristics of those flat connections.
In the present letter an explicit description of 
such flat connection for the basic case of $Spin(7)$ is given.
\end{abstract}

Recently the long standing paradox with counting the Witten 
index in $N=1$ supersymmetric gauge theory has been resolved 
\cite{Witnew}. The essence of the paradox was that different ways
of computing the Witten index gave different results for the higher
orthogonal ($SO(N),\;N \geq 7$) and the exceptional gauge groups.

The first way was to put the gauge theory into a finite
spacial box and to count
the number of supersymmetric vacuum states \cite{IWit} which resulted in
${\rm Tr} (-1)^F = r + 1$ where $r$ is the rank 
of the gauge group.  For higher orthogonal and exceptional groups, 
this result disagrees with the one based on 
counting of gluino zero modes in the 
instanton background and also on the analysis of weakly coupled theories
with additional matter super-multiplets \cite{IWit,IDyn}.
\begin{equation}
\label{IW}
{\rm Tr} (-1)^F \ =\ h^\lor\ ,
\end{equation}
where $h^\lor$ is the dual Coxeter number of the group (see e.g. \cite{K}, 
Chapt. 6; it coincides with the Casimir $T^aT^a$ in the adjoint 
representation when a proper normalization is chosen.).
For $SO(N \geq 7)$, $h^\lor = N-2 >  r+1$.
Also for exceptional groups $G_2, F_4, E_{6,7,8}$, the index (\ref{IW}) is 
larger  than the Witten's original estimate. 

In \cite{Witnew} Witten has found a flaw in his original arguments and 
shown that, for $SO(N \geq 7)$, vacuum moduli space is richer than it was 
thought before so that the {\it total} number  of quantum vacua is $N-2$ in 
accordance with the result (\ref{IW}).
The derivation in \cite{IWit} was based on the assumption that 
a flat connection on 3d torus $T^{3}$ can be gauged 
to a Cartan sub-algebra of the corresponding Lie algebra. This assumption 
seems to be quite natural since for a connected, simply connected
gauge group a connection on $T^{3}$ is flat when and only when
it has commuting holonomies ${\Omega}_{j},\;j=1,2,3$
over three independent nontrivial cycles of 
$T^{3}$, and it is very natural to expect that commuting holonomies
are representable as exponentials of commuting Lie algebra elements. 
Nevertheless, it is not true. For  higher orthogonal and exceptional 
gauge groups there are triples of commuting holonomies
which cannot be represented as exponentials of a Cartan sub-algebra
elements. This fact is in the heart of resolution of the paradox in  
\cite{Witnew}. 

Interestingly, existence of such triples has been known 
to topologists for a long (see, e.g. \cite{serr,SS}, where examples
of such triples were constructed). 

After the Witten's
work \cite{Witnew} (see also its interpretation for pedestrians in
\cite{KRS}) a new interest to such triples has arisen. 
Moduli spaces of such triples (that is, additional components
of moduli spaces of flat connections on $T^{3}$
for exceptional gauge group) have been described in 
\cite{KS} and in \cite{Ker}. Later these results have been 
re-derived and extended in some respects in \cite{Borel}. 

The purpose of this letter is to explicitely describe flat connections
corresponding to such triples (actually, only the basic case
of $Spin(7)$ is considered here; the description
is likely generalizable to other cases). Although the explicit description
is not needed in the problem of computing the Witten index,
it may be useful elsewhere, in particular, in answering the question whether
the existence of the new components of vacua moduli space affects only
the Witten index or also some other observables in supersymmetric 
gauge theories with orthogonal or exceptional gauge groups.
It may also be interesting {\it per ce}.

For $Spin(7)$ group there is a unique (up to conjugation)
nontrivial triple. It can be chosen in the form \cite{SS}, \cite{Witnew}
\newpage
\begin{eqnarray}
\label{triple}
\Omega_{1}=\gamma_{1234}\nonumber\\
\Omega_{2}=\gamma_{1256}\\
\Omega_{3}=\gamma_{1357}\nonumber,
\end{eqnarray}
where and in what follows we use the notation
\begin{equation}
\gamma_{ijkl \ldots }=\gamma_{i}\gamma_{j}\gamma_{k}\gamma_{l} \ldots
\end{equation}
and ${\gamma}_{i},\;i=1, \ldots 7$ stand for the gamma-matrices.
$\Omega$'s in Eq.(\ref{triple}) mutually commute and cannot be conjugated to
the Cartan torus (see, e.g. \cite{KRS}). 

Let $x,y,z$ be coordinates on the cube in $R^{3}$ which
gives $T^{3}$ upon identification $x \sim x+1$, $y \sim y+1$,
$z \sim z+1 $. We would like to explicitely describe a flat 
$Spin(7)$ connection 
$A_{i},\; i=1,2,3$
on $T^{3}$ which has holonomies ${\Omega}_{i},\;i=1,2,3$ from
Eq.(\ref{triple}), that is
\begin{eqnarray}
\label{holonomies}
\Omega_{1}=Pexp \left (\int_{0}^{1}A_{1}(x,0,0)dx \right)\nonumber\\
\Omega_{2}=Pexp \left (\int_{0}^{1}A_{2}(0,y,0)dy \right) \\
\Omega_{3}=Pexp \left (\int_{0}^{1}A_{3}(0,0,z)dz \right)\nonumber.
\end{eqnarray}
It will be represented in the form
\begin{equation}
\label{flat}
A_{i}(x,y,z)=g(x,y,z)^{-1} \frac{\partial}{\partial x^{i}}
g(x,y,z)
\end{equation}
where $g(x,y,z)$ takes values in the group and has the following
properties:
\begin{eqnarray}
\label{property}
g(1,y,z)=\Omega_{1}g(0,y,z)\nonumber\\
g(x,1,z)=\Omega_{2}g(x,0,z)\\
g(x,y,1)=\Omega_{3}g(x,y,0)\nonumber.
\end{eqnarray}
with $\Omega$'s from Eq.(\ref{triple}).

Obviously, with $g(x,y,z)$ obeying Eq.(\ref{property})
the flat connection $A_{i}$ from Eq.(\ref{flat}) is periodical
on $T^{3}$ and has the appropriate holonomies. 

Now introduce the following definitions:
\begin{eqnarray}
\label{edge}
g_{1}=e^{\frac{\pi}{2}x(\gamma_{12}+\gamma_{34})}\nonumber\\
g_{2}=e^{\frac{\pi}{2}y(\gamma_{15}-\gamma_{26})}\\
g_{3}=e^{\frac{\pi}{2}z(\gamma_{13}+\gamma_{57})}\nonumber.
\end{eqnarray}
$g_{i}$ Eq.(\ref{edge}) produces the monodromy $\Omega_{i}$ when the 
corresponding coordinate changes by 1. 
The specific choice of $g_{i}$'s is related to the following property:
\begin{equation}
\label{prop1}
[g_{i},\Omega_{i+1}]=0
\end{equation}
By explicit computation one verifies that
\begin{equation}
\label{prop2}
g_{i} \Omega_{i-1}=\Omega_{i-1}g_{i}^{-1}
\end{equation}

Introduce also $\tilde{g}_{3}$ such that it commutes with $g_{2}$
(and, consequently, with $\Omega_{2}$),
\begin{equation}
\label{tilde}
\tilde{g}_{3}=e^{\frac{\pi}{2}z(\gamma_{15}-\gamma_{37})}.
\end{equation}
$\tilde{g}_{3}$ as well as $g_{3}$ produces $\Omega_{3}$ 
when the corresponding coordinate ($z$) changes by 1.

One can also verify that
\begin{equation}
\label{prop3}
\tilde{g}_{3} \Omega_{1}=\Omega_{1}(\tilde{g}_{3})^{-1}.
\end{equation}

Using all these properties, one can see that the following
$g(x,y,z)$
\begin{equation}
\label{solution}
g(x,y,z)=g_{1}g_{2}g_{1}^{-1}g_{3} \tilde{g}_{3}^{-1}
g_{1}g_{2}^{-1}g_{1}^{-1}\tilde{g}_{3}g_{1}g_{2}
\end{equation}
obeys the key  property Eq.(\ref{property}),
and hence the corresponding flat connection, Eq.(\ref{flat}),
is periodic and has the monodromies $\Omega$'s.

I would like to thank A.Gorsky and A.Rosly for
discussions and A.Smilga for discussions and for convincing
me to publish this letter. This work was partially supported 
by INTAS-96-0482.

\end{document}